\pdfoutput=1
\documentclass[superscriptaddress,aps,twocolumn,floatfix,prl,showpacs]{revtex4-1}
\usepackage{graphicx,amsmath,bbm}

\newcommand{\ket}[1]{{| #1 \rangle}}
\newcommand{\bra}[1]{{\langle #1 |}}

\def\braket#1{\mathinner{\langle{#1}\rangle}}
\def\bra#1{\left\langle#1\right|}
\def\ket#1{\left|#1\right\rangle}

\begin{document}

\title{Pauli spin blockade and the ultrasmall magnetic field effect}
\date{\today}
\author{Jeroen Danon}
\affiliation{Niels Bohr International Academy, Niels Bohr
Institute, University of Copenhagen, Blegdamsvej 17, 2100
Copenhagen, Denmark}
\author{Xuhui Wang}
\author{Aur\'{e}lien Manchon}
\affiliation{King Abdullah University of Science and Technology
(KAUST), Physical Science and Engineering Division, Thuwal
23955-6900, Saudi Arabia}

\begin{abstract}
Based on the spin-blockade model for organic magnetoresistance we
present an analytic expression for the polaron-bipolaron transition rate,
taking into account the effective nuclear
fields on the sites. We reveal the physics producing qualitatively different
magnetoconductance line shapes as well as the ultrasmall magnetic
field effect, and we study the role of the ratio between the
intersite hopping rate and the typical magnitude of the nuclear
fields.
Our findings are in agreement with recent experiments and numerical simulations.
\end{abstract}
 \maketitle

The discovery some ten years ago of spin injection in organic semiconductors \cite{Dediu2002181} and a giant magnetoresistance in organic spin valves \cite{kalinowski,xiong:nature} triggered the birth of the thriving field of organic spintronics \cite{orgspinrev}, which offers interesting new physics and the potential of industrial applications.
An exciting phenomenon in this field is a large (up to 20\%) magnetoresistance observed in different organic materials \cite{francis-njp-2004,PhysRevB.72.205202,PhysRevLett.99.257201}, usually at small magnetic fields (1--10 mT) but sometimes at larger fields (10--100~mT), and persisting up to room temperature.
Since its discovery in 2004, different explanations for this organic magnetoresistance (OMAR) have been proposed: For bipolar devices it was suggested that spin-dependent electron-hole recombination and dissociation rates could be responsible \cite{Prigodin2006757,PhysRevB.76.235202}, whereas a model based on nuclear-field-mediated bipolaron formation could explain OMAR in both bipolar and unipolar devices \cite{PhysRevLett.99.216801,wagemansjap,PhysRevB.84.075204}.

More recently an organic magnetoresistive effect on an even smaller field scale (0.1--1 mT) has been observed in unipolar as well as bipolar devices \cite{PhysRevLett.105.166804}. This ultrasmall magnetic field effect (USMFE) is manifested by a sign reversal of the magnetoconductance (MC) at very small fields, creating two small peaks(dips) around zero field for devices with a negative(positive) MC.
Experimental results seem to indicate that the typical field magnitude on which the USMFE is observed scales with the width of the MC curve when different materials are investigated \cite{PhysRevLett.105.166804}.
An explanation for the effect was suggested in terms of enhanced singlet-triplet mixing close to the crossings of the hyperfine sublevels of pairs of charge carriers (polarons) coupled to single nuclear spins \cite{nguyennatmat}. This explanation is still under debate, mainly because it is expected that a single polaron in reality couples to many nuclear spins \cite{McCamey,bobbertnatmat,schultenspin}. Numerical simulations based on a semiclassical model (where the coupling to an ensemble of nuclear spins is treated as an effective magnetic field) also reproduce the USMFE \cite{PhysRevLett.106.197402} and thus invite to seek for an explanation along semiclassical lines.

Here, we study the OMAR line shape as it naturally
emerges from the spin blockade model of Ref.\
\cite{PhysRevLett.99.216801}. We present an analytic expression for
the charge current through a polaron-bipolaron link for a given realization of the
nuclear fields. Our results reproduce the USMFE and the different
line widths as observed in experiment and in numerical calculations based on the same semiclassical
approach~\cite{PhysRevLett.106.197402}, and from our analytic insight we can identify the underlying physical
mechanisms.
We note that many interesting aspects of spin-blockade physics have already 
been investigated in the seemingly foreign field of spin qubits hosted in
semiconductor quantum dots~\cite{ono:science,jouravlev:prl}, where
spin blockade is commonly used as a tool for single-qubit readout
\cite{frank:nature,reillyt2}. Indeed, the physics of the polaron
spin blockade model for OMAR is very similar to that governing the electron transport through
a double quantum dot in the spin-blockade regime~\cite{jouravlev:prl}. Our investigation thus builds on the
theoretical framework of Ref.\ \cite{jouravlev:prl}, and our explanation of the USMFE relies on a
subtlety which was not addressed in Ref.\ \cite{jouravlev:prl}.

Let us first briefly review the bipolaron model for OMAR presented in 
Ref.\ \cite{PhysRevLett.99.216801}. Electric current flows through the
organic material as polarons hop between different localized
molecular sites. 
Typically, the sites participating in transport do not
form a regular lattice
and all have a random energy offset with a distribution width $\sigma$ of 0.1--0.2 eV~\cite{PhysRevLett.99.216801}.
Sites with a relatively large negative energy offset are likely to trap a polaron for a long time, but since the on-site polaron-polaron repulsion is typically of the same order of magnitude as $\sigma$, such occupied sites
can often still take part in transport by temporarily hosting a pair of polarons, i.e.\ a bipolaron.

\begin{figure}[t]
\begin{center}
\includegraphics{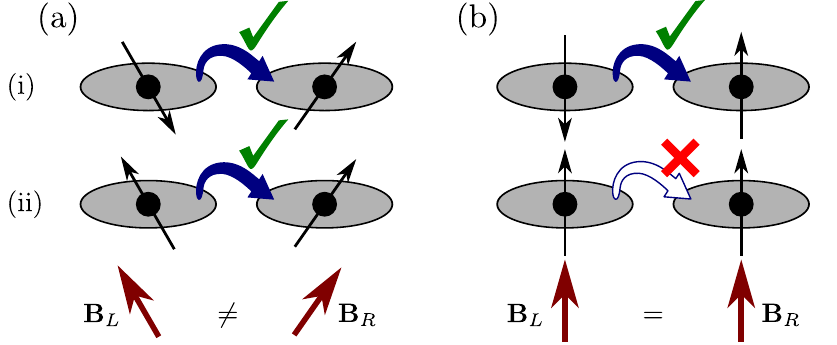}
\caption{(color online) When a site has a relatively large negative energy offset and already
contains a polaron (the right site in the pictures), charge transport through this site relies on
the formation of a bipolaron (blue arrows). Large on-site exchange effects
dictate this bipolaron to be a spin-singlet, which leads to spin-blockade physics.}\label{fig:dqd}
\end{center}
\end{figure}

Due to a relatively large orbital level spacing, most energetically accessible bipolaron states are spin singlets. This makes the polaron-bipolaron transition spin selective, ultimately leading to OMAR.
The mechanism can be understood from Fig.~\ref{fig:dqd}, where we focus on a single polaron-bipolaron transition. We assume that the spins of the two encountering polarons are random and for simplicity we describe the problem in the basis of spin eigenstates quantized along the direction of the local magnetic fields ${\bf B}_{L,R}$. 
Two possible initial spin states are depicted: (i) the left spin
antiparallel and the right spin parallel to the local field, and
(ii) both spins parallel. In the absence of an external field, the
magnetic fields at the two sites are the local random effective
nuclear fields [Fig.~\ref{fig:dqd}(a)]. Generally all initial
states can then transition to a spin-singlet bipolaron and current
runs through the system. If, on the other hand, a magnetic field
much larger than the typical nuclear fields is applied, then ${\bf
B}_L$ and ${\bf B}_R$ are (almost) parallel [Fig.~\ref{fig:dqd}(b)].
In this case, situation (ii) is a spin triplet, out of which a
bipolaron cannot be formed: the current is blocked.

We thus see that a simple two-site picture is able to explain the essentials
of OMAR. Of course, in experiment there are many possible paths
for charge carriers through the material and not all of them
contain bipolaron sites. The visibility of all effects of spin
blockade will thus be reduced, but the characteristic features
survive~\cite{PhysRevLett.99.216801}.

In this work, we will focus on the physics of a \mbox{\it single}
polaron-bipolaron transition and its MC line shape. To describe
the transition, we use five states: the four possible initial spin
states of the polaron pair (both sites hosting one polaron), one
spin-singlet state $\ket{S}$ and three spin-triplet states
$\ket{T_0}$ and $\ket{T_\pm}$, and the spin-singlet bipolaron
state $\ket{S_{\rm b}}$. The Hamiltonian we use to describe the
coherent dynamics of these states reads~\cite{jouravlev:prl}
\begin{equation}
\hat H = \left( \begin{array}{ccccc}
B_s^z & B_s^- & 0 & -B_a^- & 0 \\
B_s^+ & 0 & B_s^- & B_a^z & 0 \\
0 & B_s^+ & -B_s^z & B_a^+ & 0 \\
-B_a^+ & B_a^z & B_a^- & 0 & t \\
 0 & 0 & 0 & t & -\Delta \\
\end{array}\right),
\label{eq:ham}
\end{equation}
written in the basis $\{
\ket{T_+},\ket{T_0},\ket{T_-},\ket{S},\ket{S_{\rm b}} \}$. This
Hamiltonian includes a coupling energy $t$ between the two
singlets (which enables polaron hopping) and the relative energy
offset (detuning) $\Delta$ of the bipolaron state, typically $\Delta \sim \sigma$. The effect of
the local magnetic fields ${\bf B}_{L,R}$ is expressed in terms of
the sum and difference fields ${\bf B}_s = \tfrac{1}{2}({\bf
B}_L+{\bf B}_R)$ and ${\bf B}_a = \tfrac{1}{2}({\bf B}_L-{\bf
B}_R)$, and we use the notation $B_{s(a)}^\pm =
\tfrac{1}{\sqrt{2}}(B_{s(a)}^x \pm i B_{s(a)}^y)$.
Note that we have set $g\mu_{\rm B} = 1$ for convenience.

As pointed out in Ref.~\cite{jouravlev:prl}, we can deduce already from
Eq.~(\ref{eq:ham}) that there exist in the space of $({\bf
B}_L,{\bf B}_R)$ so-called ``stopping points'' where the current is blocked.
To see this, we take the spin quantization axis to point along ${\bf B}_s$, which amounts to setting
$B_s^\pm \to 0$ in (\ref{eq:ham}). Then we find that current vanishes when ${\bf B}_a
\parallel {\bf B}_s$ or ${\bf B}_a \perp {\bf B}_s$, since at
these points one or more of the triplet states are not coupled to $\ket{S}$.
The sum and difference fields ${\bf B}_{s,a}$ both contain a
contribution from the effective nuclear fields ${\bf K}_{L,R}$ on
the two sites, whereas the external field ${\bf B}_{\rm ext}$ only
adds to the sum field: ${\bf B}_s = {\bf K}_s + B_{\rm ext}\hat z$
and ${\bf B}_a = {\bf K}_a$. For a given random realization of
${\bf K}_{L,R}$ one can thus always find a field $B_{\rm ext}$ for
which ${\bf B}_a \perp {\bf B}_s$, and a sweep of $B_{\rm ext}$
for a {\it fixed} ${\bf K}_{L,R}$ will always exhibit a stopping
point where the current vanishes. The position of this
stopping point is determined by the relative orientation of ${\bf
K}_{s}$ and ${\bf K}_{a}$ and is thus random. In an experiment one
usually sweeps $B_{\rm ext}$ so slowly that at each measurement
many configurations of the fields ${\bf K}_{L,R}$ are probed. As a
result the stopping points are averaged out and one finds a smooth
MC curve \cite{jouravlev:prl}.

However, this is not the full story. A subtlety, not discussed in
Ref.\ \cite{jouravlev:prl}, is that there exists one more stopping
point \cite{omar.note1}: When ${\bf B}_s = 0$ the triplet subspace
in the Hamiltonian is degenerate and Eq.~(\ref{eq:ham}) can be
equivalently written in terms of one coupled triplet state
\begin{equation*}
 \ket{T_m} = \frac{-B_a^-\ket{T_+} + B_a^z\ket{T_0} + B_a^+\ket{T_-}}{|{\bf B}_a|},
\end{equation*}
and two orthogonal triplet states $\ket{T_{1}}$ and $\ket{T_{2}}$
which have $\braket{T_{1,2}|\hat H|S} = 0$ and are thus blocked.
Why would we bother? We argued above that stopping points occur at
random positions, leaving no trace after averaging over ${\bf
K}_{L,R}$. This new stopping point however, is fundamentally different from
the ones discussed above: It suppresses current close to where
${\bf B}_{\rm ext} = -{\bf K}_s$, which is {\it always} in the
vicinity of $B_{\rm ext} = 0$. It is therefore possible that
after averaging over ${\bf K}_{L,R}$ this new stopping point
leaves a trace in the MC curve: A small dip around zero
field, like the USMFE.

Let us now explicitly calculate the current as governed by this polaron-bipolaron
transition. To describe charge transport, we write a
time-evolution equation for the $5\times 5$ density matrix of the
system. To the coherent evolution dictated by
$\hat H$ we add incoherent rates describing dissociation of
the bipolaron to the environment and hopping of a new polaron onto
the empty left site. This yields (where we have set $\hbar$ to $1$)
\begin{equation}
\frac{\partial \hat \rho}{\partial t} = -i [\hat H,\hat \rho]
- \frac{\Gamma}{2}\{\hat P_{\rm b}, \hat \rho \}
+ \frac{\Gamma}{4}\rho_{{\rm b},{\rm b}}(1-\hat P_{\rm b}) \hat{\mathbbm{1}},
\end{equation}
where $\Gamma$ is the rate of bipolaron dissociation to the
environment, $\hat P_{\rm b} = \ket{S_{\rm b}}\bra{S_{\rm b}}$ is
the projection operator onto the bipolaron state, and
$\hat{\mathbbm{1}}$ is the identity matrix. In writing so we
assumed for simplicity that refilling of the left site takes place
immediately after dissociation of the bipolaron. If this is not
the case, the prefactor for the current changes but the MC
characteristics stay the same.

We add the normalization condition ${\rm Tr}[\hat\rho] = 1$ to the
set of equations and then solve $\partial_t\hat\rho^{({\rm eq})} =
0$ to find the stationary density matrix. The charge current is then given by $I = e\Gamma \rho^{({\rm eq})}_{{\rm
b},{\rm b}}$ and can be found explicitly. We assume for
convenience that $\Gamma \gg t, B_s, B_a, \Delta$ is the largest
energy scale in the problem \cite{PhysRevB.84.075204}, and then
find
\begin{align}
I = e\Gamma_s \frac{4 x^2 \sin^2\phi}{x^4+ax^2+1}, \label{eq:curr}
\end{align}
in terms of $x \equiv B_s/B_a$.  Here $\Gamma_s \equiv t^2/\Gamma$
is the singlet-singlet hopping rate from the left to the right
site and $\phi$ is the angle between ${\bf B}_s$ and ${\bf
B}_a$~\cite{omar.note2}. We also used
\begin{align}
a = \frac{\Gamma_s^2}{B_a^2} \left(3+\frac{1}{\cos^{2}\phi}\right) -2\cos 2\phi.
\label{eq:a}
\end{align}
We see that all stopping points predicted above are indeed
reflected in (\ref{eq:curr}): At $B_{s}=0$ we have $x=0$ which
yields $I=0$, and ${\bf B}_a \parallel {\bf B}_s$ or ${\bf B}_a
\perp {\bf B}_s$ corresponds to $\phi = 0,\pi$ or $\phi = \pi/2$
respectively, both also giving $I=0$.

Equation (\ref{eq:curr}) is the most important analytic result of
our work. It gives the current for one single realization of ${\bf
K}_L$, ${\bf K}_R$, and $B_{\rm ext}$. The MC measured in
experiment is found by averaging (\ref{eq:curr}) over the random
nuclear fields. In contrast to the analytic results presented in
\cite{jouravlev:prl}, our result is valid for arbitrary $\Gamma_s$
and not only for limiting cases. One word of caution is required
here concerning the interpretation of (\ref{eq:curr}): If one
wants to plot $I(B_{\rm ext})$ for a single realization of ${\bf
K}_{L,R}$, one should not only use $B_s = |{\bf K}_s + B_{\rm
ext}\hat z|$ in (\ref{eq:curr}) but also implement the dependence
of $\phi$ on $B_{\rm ext}$ implied by $\cos\phi = ({\bf
B}_s\cdot{\bf B}_a)/B_sB_a$.

\begin{figure}[t]
\begin{center}
\includegraphics{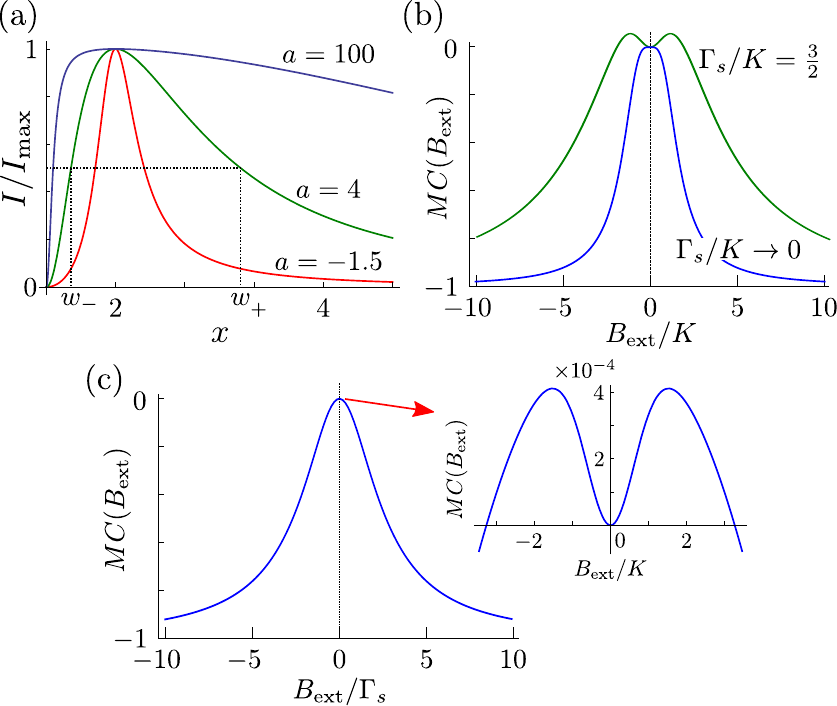}
\caption{(color online) (a) The current given by (\ref{eq:curr}) as a function of $x=B_s/B_a$ for fixed $B_a$ and $\phi$, evaluated for different parameters $a$. (b,c) The averaged MC. (b) Blue trace: $\Gamma_s/K\ll 1$. The field $B_{\rm ext}$ is plotted in units of $K = \langle {K}^2_{L,R}\rangle^{1/2}$. The peak of this curve is flat~\cite{jouravlev:prl}. Green trace: $\Gamma_s/K = \tfrac{3}{2}$. (c) $\Gamma_s/K = 50$. Now $B_{\rm ext}$ is plotted in units of $\Gamma_s$. (inset) The range where $B_{\rm ext}\sim K$.}\label{fig:plot2}
\end{center}
\end{figure}
Let us now investigate Eq.~(\ref{eq:curr}) and see what we can
infer  about the line shape of the predicted MC curve. We always
have $a > -2$, which ensures that $I\geq0$ everywhere. The current
vanishes for $x=0$ or $x\to \infty$, and in the range $x \in
[0,\infty]$ we have a single maximum at $x=1$ where the current is
$I_{\rm max} = 4e\Gamma_s\sin^2\phi/(a+2)$. In Fig.~\ref{fig:plot2}(a)
we plot the expression given in Eq.~(\ref{eq:curr}) for different
$a$. The FWHMs $w_-$ of the dip around $x=0$ and $w_+$ of the
overall peak structure (as indicated in the plot for $a=4$) are
found to be $w^2_\pm = 2+\tfrac{1}{2}a \pm
\sqrt{3+2a+\tfrac{1}{4}a^2}$.

We see from Eq.~(\ref{eq:a}) that an important parameter is
$\Gamma_s/K$,  the ratio of the intersite hopping rate and the
typical magnitude of the nuclear fields $K$, typically $\sim 0.1~\mu$eV~\cite{McCamey}.
We will thus now investigate the cases of small and large $\Gamma_s/K$.

In the limit of $\Gamma_s/K \ll 1$ we can write
\begin{equation}
I \approx e\Gamma_s \frac{4 x^2 \sin^2\phi}{x^4-2x^2\cos 2\phi +1} = \Gamma_s ({\bf n}_L\times{\bf n}_R)^2,
\label{eq:se}
\end{equation}
where we used the unit vectors ${\bf n}_{L,R} = {\bf
B}_{L,R}/B_{L,R}$.  As it should, this result coincides with that
of Ref.~\cite{jouravlev:prl} in the same limit: There are no
intersite exchange effects and the situation is exactly like the
picture of Fig.~\ref{fig:dqd} where the current only depends
on the relative orientation of ${\bf B}_L$ and ${\bf B}_R$. As was
shown in \cite{jouravlev:prl}, Eq.~(\ref{eq:se}) can be averaged
analytically over random ${\bf K}_{L,R}$ taken from a normal
distribution, yielding a MC curve with a flat peak at $B_{\rm ext}
= 0$, a maximum of $\langle I\rangle_{\rm max} \sim e\Gamma_s$, and
a line width of $\sim K$. Indeed, for all $a \in [-2,2]$ we find
that $w_+ \sim 1$, so for any $\phi$ the current is suppressed
when $x\gtrsim 1$. In Fig.~\ref{fig:plot2}(b) (blue trace) we plot the resulting
MC line shape, where we defined $MC(B_{\rm ext}) = [I(B_{\rm
ext})-I(0)]/I(0)$.

In the opposite limit of $\Gamma_s/K \gg 1$ we have $a \approx
(\Gamma_s/B_a)^2(3+\cos^{-2}\phi) \gg 1$. We can already see from
the properties of Eq.~(\ref{eq:curr}) that in this case $\langle
I\rangle_{\rm max} \sim eK^2/\Gamma_s$, and that $w_+ \approx
a^{1/2} \sim \Gamma_s/K$ implies a MC line width of $\sim
\Gamma_s$. Indeed, $\Gamma_s$ sets the level broadening of
$\ket{S}$ and as long as $B_s \lesssim \Gamma_s$ generally all
three triplet states can efficiently transition to $\ket{S}$ with
the coupling provided by ${\bf B}_a$. The width of the dip around
the stopping point at $x=0$ is $w_- \approx a^{-1/2}\sim
K/\Gamma_s$ in terms of $x$, or $\sim K^2/\Gamma_s$ in terms of
$B_s$. This energy scale can also be understood: If $B_s=0$ the
decay rate of $\ket{T_m}$ is $\Gamma_t \sim K^2/\Gamma_s$, which
is the only energy relevant in the triplet subspace. When $B_s
\gtrsim \Gamma_t$ the decay of the other two triplet states
becomes comparable to $\Gamma_t$ and the blockade is lifted.

In this limit of large $\Gamma_s/K$ the current cannot be averaged
analytically over the nuclear fields, and one has to evaluate the
integrals over the distribution of ${\bf K}_{L,R}$ numerically.
Fig.~\ref{fig:plot2}(c) shows a plot of the MC integrated over
normal distributions for all six components of ${\bf K}_L$ and
${\bf K}_R$. For all components we used a standard deviation of
$K/\sqrt{3}$ and we have set $\Gamma_s/K=50$. The resulting line
shape is Lorentzian since it is determined by the level broadening
of $\ket{S}$. Close to zero field, where $B_{\rm ext} \sim K$, we
find a very faint USMFE, as shown in the inset. When we set
$\Gamma_s/K$ even larger we find that the visibility of this USMFE
is suppressed further, ultimately reaching zero.

\begin{figure}[t]
\begin{center}
\includegraphics{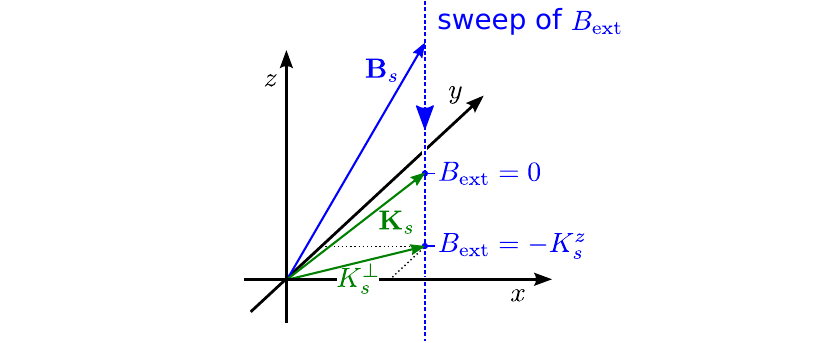}
\caption{When $B_{\rm ext}$ is swept for a given realization of ${\bf K}_{L,R}$, the field ${\bf B}_s$ (blue arrow) follows a trace like the blue dashed line. We indicated with green arrows ${\bf K}_s$ as well as $K_s^\perp$, which equals the minimum value of $B_s$.}\label{fig:dir}
\end{center}
\end{figure}
We can understand this USMFE from the expression for the current
given in Eq.~(\ref{eq:curr}). For a given realization of ${\bf
K}_{L,R}$, the current trace $I(B_{\rm ext})$ ``misses'' the
zero-field stopping point by $K_s^\perp =
\sqrt{(K_s^x)^2+(K_s^y)^2}$, as illustrated in Fig.~\ref{fig:dir}. Some realizations have $K_s^\perp
\geq B_a$ so that the current trace has a single maximum
[see Fig.~\ref{fig:plot2}(a)]. Other realizations have
$K_s^\perp < B_a$ and the current exhibits a dip at small fields,
the position of the dip at $B_{\rm ext} = -K_s^z$. For large
$\Gamma_s/K$ the dip around the zero-field stopping point becomes
narrow, of the order $\sim K^2/\Gamma_s \ll K$, and only the very
few curves of $I(B_{\rm ext})$ with $K_s^\perp \lesssim
K^2/\Gamma_s$ have an appreciable dip. This still can produce a
faint dip in the averaged current. However, the position of each
single-realization narrow dip is at $B_{\rm ext} = -K_s^z$, so
averaging over $K_s^z$ makes the averaged dip even less pronounced
and results in a dip width of $\sim K$.

The regime to look for a pronounced USMFE is thus at intermediate
$\Gamma_s/K \sim 1$. In Fig.~\ref{fig:plot2}(b) (green trace) we plot the averaged
MC for $\Gamma_s/K = 3/2$ and we see indeed a strong USMFE, its
visibility being $\sim 5$\%. This regime is optimal for the USMFE
since here the width of the zero-field dip is still $\sim K$ but
the symmetric situation where the current only depends on the
angle between ${\bf n}_L$ and ${\bf n}_R$ is significantly
perturbed. In other words, at $\Gamma_s/K\to 0$ the overall MC
line width is minimal and $\sim K$. The two USMFE ``bumps'' are
still there but are split by the same energy scale $\sim K$ and
thus appear just left and right of the top of the MC curve. In the
limit of $\Gamma_s/K = 0$ the bumps and the underlying MC curve
have exactly compatible shape line shapes, together resulting in
the characteristic flat peak. If one moves away from $\Gamma_s/K =
0$ the underlying MC line shape becomes broader, which makes the
USMFE bumps more visible. However, as soon as $\Gamma_s/K$ becomes
too large, one enters the regime discussed above, where the USMFE
disappears again. The optimal regime is thus at $\Gamma_s/K \sim
1$, in agreement with the results presented in
Fig.~\ref{fig:plot2}(b,c) as well as with previously obtained numerical results~\cite{PhysRevB.84.075204}.

To summarize, we studied the two-site spin-blockade model for OMAR and
derived an analytic expression for the polaron-bipolaron transition rate,
taking into account the local nuclear fields on the two sites. We
showed how our result reproduces different MC line widths: $\sim
K$ for slow and $\sim \Gamma_s$ for fast intersite hopping. We
also provided an explanation of the USMFE in terms of a persistent
spin blockade at the special point where the average effective field vanishes
$B_s = 0$. The USMFE as predicted here always takes place on the field scale $\sim K$, and
we explained why it is expected to be most pronounced in the
regime where $\Gamma_s \sim K$. In this regime thus both the scale
of the USMFE and the MC line width are set by $K$, the latter
however being slightly larger. This relation between the two
scales is consistent with experimental
observations~\cite{PhysRevLett.105.166804} and numerical
simulations~\cite{PhysRevB.84.075204}.

As a side remark we note here that a close inspection of the
experimental data presented in Ref.~\cite{jouravlev:prl} (the
current through a double quantum dot) also seems to reveal a faint
USMFE. The data were fitted to the flat-peak curve since the
system was assumed to be in the inelastic tunneling regime. In
reality the coupling is however never perfectly inelastic, and a
faint trace of the USMFE could be left. Due to its tunability, a
double quantum dot might in fact be the best system to
experimentally explore USMFE in more detail.

We acknowledge helpful feedback from M.~S.~Rudner and P.~A.~Bobbert.


\begin{thebibliography}{10}

\bibitem{Dediu2002181}
V.~Dediu, M.~Murgia, F.~Matacotta, C.~Taliani, and S.~Barbanera,
Solid State Commun. \textbf{122}, 181  (2002).

\bibitem{kalinowski}
J.~Kalinowski, M.~Cocchi, D.~Virgili, P.~{Di~Marco}, V.~Fattori
Chem. Phys. Lett. \textbf{380}, 710 (2003).

\bibitem{xiong:nature}
Z.~H. Xiong, D.~Wu, Z.~Valy~Vardeny, and J.~Shi, Nature (London)
\textbf{427}, 821 (2004).

\bibitem{orgspinrev}
V.~A. Dediu, L.~E. Hueso, I.~Bergenti, and C.~Taliani, Nat. Mater.
\textbf{8}, 707 (2009).

\bibitem{francis-njp-2004}
T.~L. Francis, {\" O}.~Mermer, G.~Veeraraghavan, and
M.~Wohlgenannt,  New Journal of Physics \textbf{6}, 185 (2004).

\bibitem{PhysRevB.72.205202}
{\" O}.~Mermer, G.~Veeraraghavan, T.~L. Francis, Y.~Sheng, D.~T.
Nguyen, M.~Wohlgenannt, A.~K\"ohler, M.~K. Al-Suti, and M.~S.
Khan, Phys. Rev. B \textbf{72}, 205202 (2005).

\bibitem{PhysRevLett.99.257201}
F.~L. Bloom, W.~Wagemans, M.~Kemerink, and B.~Koopmans, Phys. Rev.
Lett. \textbf{99}, 257201 (2007).

\bibitem{Prigodin2006757}
V.~Prigodin, J.~Bergeson, D.~Lincoln, and A.~Epstein, Synthetic
Metals \textbf{156}, 757  (2006).

\bibitem{PhysRevB.76.235202}
P.~Desai, P.~Shakya, T.~Kreouzis, and W.~P. Gillin, Phys. Rev. B \textbf{76}, 235202 (2007).

\bibitem{PhysRevLett.99.216801}
P.~A. Bobbert, T.~D. Nguyen, F.~W.~A. van Oost, B.~Koopmans, and
M.~Wohlgenannt, Phys. Rev. Lett. \textbf{99}, 216801 (2007).

\bibitem{wagemansjap}
W.~Wagemans, F.~L. Bloom, P.~A. Bobbert, M.~Wohlgenannt, and
B.~Koopmans, J. Appl. Phys. \textbf{103}, 07F303 (2008).

\bibitem{PhysRevB.84.075204}
A.~J. Schellekens, W.~Wagemans, S.~P. Kersten, P.~A. Bobbert, and
B.~Koopmans, Phys. Rev. B \textbf{84}, 075204 (2011).

\bibitem{PhysRevLett.105.166804}
T.~D. Nguyen, B.~R. Gautam, E.~Ehrenfreund, and Z.~V. Vardeny,
Phys. Rev. Lett. \textbf{105}, 166804 (2010).

\bibitem{nguyennatmat}
T.~D. Nguyen, G.~Hukic-Markosian, F.~Wang, L.~Wojcik, X.-G. Li,
E.~Ehrenfreund, and Z.~V. Vardeny, Nat. Mater. \textbf{9}, 345
(2010).

\bibitem{McCamey}
D.~R. {McCamey}, K.~J. {van Schooten}, W.~J. Baker, {S.-Y.} Lee, {S.-Y.} Paik, J.~M. Lupton, and C. Boehme
Phys. Rev. Lett. \textbf{104}, 017601 (2010).

\bibitem{bobbertnatmat}
P.~A. Bobbert, Nat. Mater. \textbf{9}, 288 (2010).

\bibitem{schultenspin}
K.~Schulten and P.~G. Wolynes, J. Chem. Phys. \textbf{68}, 3292 (1978).

\bibitem{PhysRevLett.106.197402}
S.~P. Kersten, A.~J. Schellekens, B.~Koopmans,  and P.~A. Bobbert,
Phys. Rev. Lett. \textbf{106}, 197402 (2011).

\bibitem{ono:science}
K.~Ono, D.~G. Austing, Y.~Tokura, and S.~Tarucha, Science \textbf{297}, 1313 (2002).

\bibitem{jouravlev:prl}
O.~N. Jouravlev and Y.~V. Nazarov, Phys. Rev. Lett. \textbf{96}, 176804 (2006).

\bibitem{frank:nature}
F.~H.~L. Koppens, C.~Buizert, K.~J. Tielrooij, I.~T. Vink, K.~C.
Nowack, T.~Meunier, L.~P. Kouwenhoven, and L.~M.~K. Vandersypen,
Nature \textbf{442},
  766 (2006).

\bibitem{reillyt2}
D.~J. Reilly, J.~M. Taylor, J.~R. Petta, C.~M. Marcus, M.~P.
Hanson, and A.~C. Gossard, Science \textbf{321}, 817 (2008).

\bibitem{omar.note1}
One could argue that ${\bf B}_a = 0$ is also a stopping point,
but it is not of any interest since it occurs independently from
$B_{\rm ext}$ and therefore leaves no traces in the MC curve. In
that sense it is also not really a stopping {\it point}. Besides,
it is captured by the model in \cite{jouravlev:prl} since it is
equivalent to having ${\bf B}_a \parallel {\bf B}_s$ {\it and}
${\bf B}_a \perp {\bf B}_s$.

\bibitem{omar.note2}
If one would relax the assumption $\Gamma \gg \Delta$, one finds
the same expression but with $\Gamma_s \to t^2/\sqrt{\Gamma^2 +
4\Delta^2}$ and an extra prefactor $\Gamma / \sqrt{\Gamma^2 +
4\Delta^2}$. A finite detuning $\Delta$ thus merely leads to a
suppression of the current as well as a smaller effective hopping
rate.

\end{thebibliography}
\end{document}